\documentclass[conference,letterpaper]{IEEEtran}
\usepackage[utf8]{inputenc} 
\usepackage[T1]{fontenc}
\usepackage{amsfonts}
\usepackage{algorithmicx}
\usepackage{algorithm}
\usepackage{algpseudocode}
\usepackage{array}
\usepackage{ifthen}
\usepackage[caption=false, font=normalsize, labelfont=sf, textfont=sf]{subfig}
\usepackage{textcomp}
\usepackage{stfloats}
\usepackage{url}
\usepackage{verbatim}
\usepackage{graphicx}
\usepackage{cite}
\usepackage[cmex10]{amsmath}
\begin{document}

\title{DALC: Distributed Arithmetic Coding Aided by Linear Codes}

\author{%
\IEEEauthorblockN{Junwei Zhou, HaoYun Xiao and Jianwen Xiang}
\IEEEauthorblockA{the School of Computer Science and Artificial Intelligence \\
                    Wuhan University of Technology\\
                    Wuhan\\
                  Email: \{junweizhou,346765, jwxiang\}@whut.edu.cn}
  \and
  \IEEEauthorblockN{Qiuzhen Lin}
  \IEEEauthorblockA{College of Computer Science and Software Engineering\\ 
                    Shenzhen University University \\
                    Shenzhen University\\
                    Email: qiuzhlin@szu.edu.cn
                    }
}

\markboth{ISIT2025}%
{Junwei \MakeLowercase{\textit{et al. }}: Distributed Arithmetic Coding Aided by Linear Codes}

\maketitle
\begin{abstract}
Distributed Arithmetic Coding (DAC) has emerged as a feasible solution to the Slepian-Wolf problem, particularly in scenarios with non-stationary sources and for data sequences with lengths ranging from small to medium. Due to the inherent decoding ambiguity in DAC, the number of candidate paths grows exponentially with the increase in source length. To select the correct decoding path from the set of candidates, DAC decoders utilize the Maximum A Posteriori (MAP) metric to rank the decoding sequences, outputting the path with the highest MAP metric as the decoding result of the decoder. However, this method may still inadvertently output incorrect paths that have a MAP metric higher than the correct decoding path, despite not being the correct decoding path. To address the issue, we propose Distributed Arithmetic Coding Aided by Linear Codes (DALC), which employs linear codes to constrain the decoding process, thereby eliminating some incorrect paths and preserving the correct one. During the encoding phase, DALC generates the parity bits of the linear code for encoding the source data. In the decoding phase, each path in the set of candidate paths is verified in descending order according to the MAP metric until a path that meets the verification criteria is encountered, which is then outputted as the decoding result. DALC enhances the decoding performance of DAC by excluding candidate paths that do not meet the constraints imposed by linear codes. Our experimental results demonstrate that DALC reduces the Bit Error Rate(BER), with especially improvements in skewed source data scenarios.
\end{abstract}


\section{Introduction}
\IEEEPARstart{D}{istributed} Arithmetic Coding (DAC)\cite{DAC2009} is a novel coding method that address the Slepian-Wolf problem \cite{Slepian-Wolf1973} using arithmetic coding. In contrast to conventional Slepian-Wolf Coding (SWC), which is usually founded on channel coding methodologies, DAC demonstrates adaptability to sources with non-stationary behavior and superior handling capabilities for data sequences ranging from short to medium in length \cite{SeDAC2010}. 

DAC can be briefly categorized into two subtypes: Interval-Overlapping DAC (IO-DAC) \cite{IODAC2007} and Bit-Puncturing DAC (BP-DAC) \cite{BPDAC2009}. The IO-DAC approach extends and intersects the intervals associated with source symbols. The BP-DAC method involves periodically eliminating certain bits from the Arithmetic Coding (AC) bitstreams. In this paper, unless specified otherwise, DAC denotes IO-DAC. 

Since the advent of DAC, research in this area has rapidly expanded, with various studies aimed at expanding its potential applications. M. Grangetto \textit{et al}. extended DAC into rate-compatible coding \cite{Rate-compatibleDAC2008}, successfully leveraging a feedback channel between the encoder and the decoder. The time-shared Distributed Arithmetic Coding (TS-DAC) \cite{TSDAC2007} implements SWC, where the two correlated information sources, $X$ and $Y$, alternately serve as the source and side information. Lossless Adaptive-Distributed Arithmetic Coding (LADAC)\cite{LADAC2010} builds upon this foundation to achieve lossless adaptive compression without the need for known prior knowledge, enabling the encoder and decoder to operate simultaneously rather than alternately. Evidence from \cite{JZ2013} suggests that the minimum Hamming distance can be minimized to approach a value close to one. 

DAC differs from AC in its subinterval expansion, causing partial overlaps that introduce ambiguity in decoding, leading to an exponential growth in candidate decoding paths with increasing block length. Consequently, DAC decoding necessitates side information to determine the correct path. The decoding process in DAC is symbol-driven and sequential, involving a tree search where each node signifies a decoder state. DAC's inherent ambiguity means both overlapping paths must be stored if a decoding path falls into an overlapping interval, potentially resulting in multiple candidate outcomes. To improve performance, pruning is crucial in the decoding process. Existing DAC decoding often employs the $M$-algorithm, using symmetric or asymmetric sources as side information to aid pruning, though these methods may not always yield effective results.

The $M$-algorithm retains the M largest MAP estimates among candidate paths. A too-small $M$ risks excluding the correct path prematurely, while a too-large M may introduce excessive interference paths. Meanwhile, we note that a significant issue in DAC decoding is the susceptibility to errors in the tail bits. On one hand, since the encoding process accumulates symbol by symbol, the tail bits are highly sensitive to the precision of the encoding results. On the other hand, the MAP calculation involves numerous probability multiplications, where even minor computational errors can gradually amplify during cumulative multiplication, ultimately affecting the accuracy of the tail bits. To address this problem, M. Grangetto et al. \cite{DAC2009} reduced ambiguity in tail paths by configuring the last $t$ bits in a non-expanding interval state, thereby improving performance. However, we argue that this measure remains insufficient. We observe that the decoding process of DAC exhibits certain similarities with the List Decoding of Polar Codes, where CRC is employed to screen decoding paths \cite{Tal2015}. Inspired by this, we propose the Distributed Arithmetic Coding Aided by Linear Codes (DALC) method, which eliminates interfering paths using linear codes.

DALC encodes the parity check bits of a linear code into the source $X$. During decoding, the linear code's verification method checks candidate paths in descending MAP order, and the verified path is outputted as the result.

The $M$-algorithm of DALC is effectively operating on a subset $\omega$ of the original candidate path set $\Omega$. The subset $\omega$ comprises paths from $\Omega$ that have passed the linear code verification. Since the correct decoding path will not fail the linear code verification, it is ensured to be retained in $\omega$. In contrast, the primary interfering paths, those with a MAP metric higher than that of the correct decoding path, will not be included in $\omega$ if they do not pass the linear code verification. By employing this method, DALC can improves the decoding accuracy and decoding performance. Simulation experiments demonstrate that DALC outperforms existing DAC methods in terms of Bit Error Rate(BER), particularly for asymmetric sources. 

The remainder of this paper is organized as follows. Section \ref{related} introduces the related work on the optimization of the DAC decoding process. Section \ref{dalc} introduces DALC. Section \ref{result} provides the simulation results of the experiments. Finally, Section \ref{conclusion} will conclude the work. 

\section{Related Work}
\label{related}

M. Grangetto \textit{et al}.\cite{DAC2009} deduced the MAP metric $\hat{X}_{MAP}$ of the candidate path $X$ as follows: 
\begin{equation}\hat{X}_{MAP} =\prod_{i=1}^N\frac{p(x_i)p(y_i|x_i)}{p(y_i)}\end{equation}

To more effectively identify the correct decoding path from the set of candidate paths, researchers have developed various methods. In this context, the majority of research has focused on refining the MAP metric calculation for DAC. 

Y. Fang explored the codeword distribution of DAC for symmetric binary sources\cite{DistributionofDAC2009}. To deepen the theoretical understanding of DAC, two analytical tools have been introduced: the Coset Cardinality Spectrum (CCS) \cite{ImDAC2014, DACspectrum2013}, and the Hamming Distance Spectrum (HDS) \cite{HDS2016}. By incorporating CCS information into the MAP metric, Y. Fang \textit{et al}. modified the path metric of $X$ as follows:

\begin{equation}\hat{X}_{MAP} =\prod_{i=1}^N\frac{p(x_i|C_X)p(y_i|x_i)}{p(y_i)}\end{equation}

Y. Fang also validated the correctness of the theoretical analysis in \cite{TwoCSS2021}. Utilizing CCS and HDS, Y. Fang \textit{et al}. \cite{AnaTailedDAC2016} have analyzed the performance gap between Tailed DAC and Tailless DAC. To simplify computations, some studies have proposed approximation methods for the initial CCS. N. Yang \textit{et al}. utilize interpolation and bell-shaped approximations to generate simplified approximations of the initial CCS at low coding rates\cite{ADAC2023}. 

Modified Decoding Metric for Distributed Arithmetic Coding (MDAC) \cite{MDAC2020} improves the MAP metric by not considering prior information, thereby providing the correct decoding path with a competitive advantage and improving performance. The modified metric for path $X$ can be expressed as:

\begin{equation}\hat{X}_{MAP} =\prod_{i=1}^N\frac{p(y_i|x_i)}{p(y_i)}\end{equation}

Context-Based Distributed Arithmetic Coding with Sampling (CDACS) \cite{CDACS2020} enhances the computation of the MAP metric by using contextual modeling, thereby surpassing the performance of DAC. 

Additionally, there are some researches that have improved the mechanism of the $M$-algorithm.

In contrast to the existing Breadth-First Decoder (BFD), a Depth-First Decoder (DFD) \cite{DFD2020} has been proposed. The process of the DFD fully searching the DAC tree requires visiting only a small subset of nodes, effectively reducing the number of candidate decoding paths. J. Zhou \textit{et al}. employ the interval swapping technique, leveraging a priori knowledge of the source to reduce the number of candidate paths\cite{JZ2015}.

To further enhance the decoding performance of DAC, we have developed the DALC method, which refines the $M$-algorithm based on the MDAC\cite{MDAC2020}. DALC introduces an additional layer of sophistication by integrating linear code constraints into the MDAC\cite{MDAC2020} framework.

This integration is not merely an enhancement but a strategic evolution, leveraging the strength of linear codes to further refine the MAP metric. By imposing these constraints during the encoding phase, DALC ensures that only decoding paths that align with the linear code properties are considered, thereby potentially increasing the accuracy and efficiency of the decoding process beyond what MDAC\cite{MDAC2020} achieves. The DALC method, therefore, can be seen as a natural progression from MDAC\cite{MDAC2020}, offering a more nuanced approach to decoding by incorporating the benefits of linear codes.

\section{The Proposed Method}
\label{dalc}

\begin{figure}[ht]
  \centering
  \includegraphics[width=8cm, page={1}]{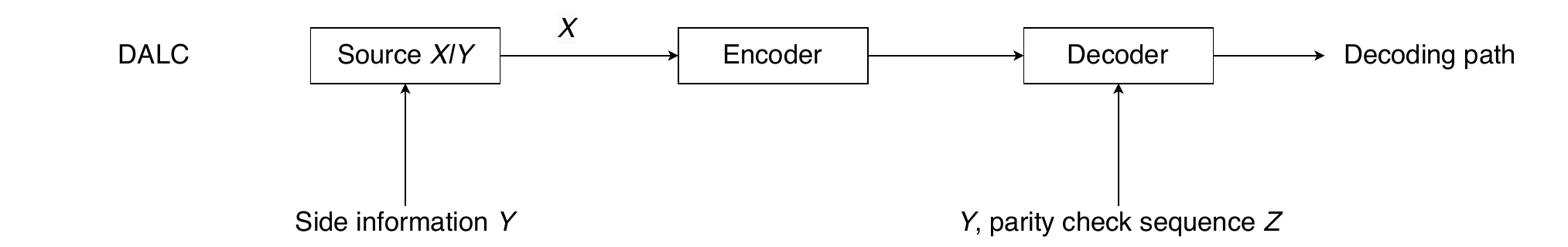}
  \caption{The framework diagram of DAC and DALC. }
  \label{fig:total}
\end{figure}

The framework diagram of DALC is shown in Fig. \ref{fig:total}. In the decoder of DALC, the parity check sequence $Z$ is employed to aid in the selection of the decoding path.

\subsection{Encoder of DALC}
Let $X = (x_0, x_1, \ldots, x_{n-1})$ be a binary symbol sequence of length $n$. The probabilities are defined as $p_0 = P (x_i = 0)$ and $p_1 = P (x_i = 1)$. 
The source $X$ is firstly encoded by a linear systematic code to produce $Z$ is represented as follows: 
\begin{equation}
Z = X\mathbf{G}^T,     
\end{equation}
where $Z = (z_0, z_1, \ldots, z_{m-1})$ represents the parity checks sequence, $\mathbf{G}$ is the generator matrix corresponding to the linear code. Subsequently, $Z$ along with $Y = (y_0, y_1, \ldots, y_{n-1})$, which is a side information binary symbol sequence of length n, is transmitted to the decoder as the side information sequence. During the encoding process of DALC, the two ranges corresponding to the symbols $1$ and $0$ are modified and expanded to $p_0'$ and $p_1'$. Specifically, $p_0' = p_{0}^{\alpha}$ and $p_1' = p_{1}^{\alpha}$, where $\alpha$ is known as the overlap factor, with $0\leq \alpha \leq 1$. Consequently, the sub-intervals mapped by 0 and 1 become $[0, p_0')$ and $[1 - p_1', 1)$ respectively. Let $[L_i, H_i)$ represent the interval output after encoding the $i$ symbols. The encoder of DALC initializes the interval to $L_0=0, H_0=1$. When $0 \leq i \leq (n-t-1)$, the $i$-th symbol is mapped into the next subinterval, that is:
\begin{equation}
\left[ L_{i+1}, H_{i+1} \right) = 
\begin{cases}
\left[ L_{i}, L_{i} + p_{0}^{\prime} (H_{i} - L_{i}) \right), & \text{if } x_{i} = 0 \\
\left[ L_{i} + (1 - p_{1}^{\prime}) (H_{i} - L_{i}), H_{i} \right), & \text{if } x_{i} = 1
\end{cases}
\end{equation}
In the last $t$ symbols, equation (5) still applies, but $ \alpha $ is fixed to 1. That is, when $n-t-1 < i \leq n-1$, in equation (5) we have $p_{0}'=p_{0}$ and $p_{1}'=p_{1}$.

After encoding $n$ symbols, $X$ can be represented by the final output interval $[L_{n}, H_{n})$. Typically, a binary number is selected from this interval to serve as the codeword $C_X$. Describing this number and the codeword requires approximately $-\log_2 (H_{n}-L_{n})$ bits. Overlapping results in a larger final interval, thus DALC generates a shorter $C_X$ than AC. 

\begin{figure}[ht]
  \centering
  \includegraphics[width=8cm, page={1}]{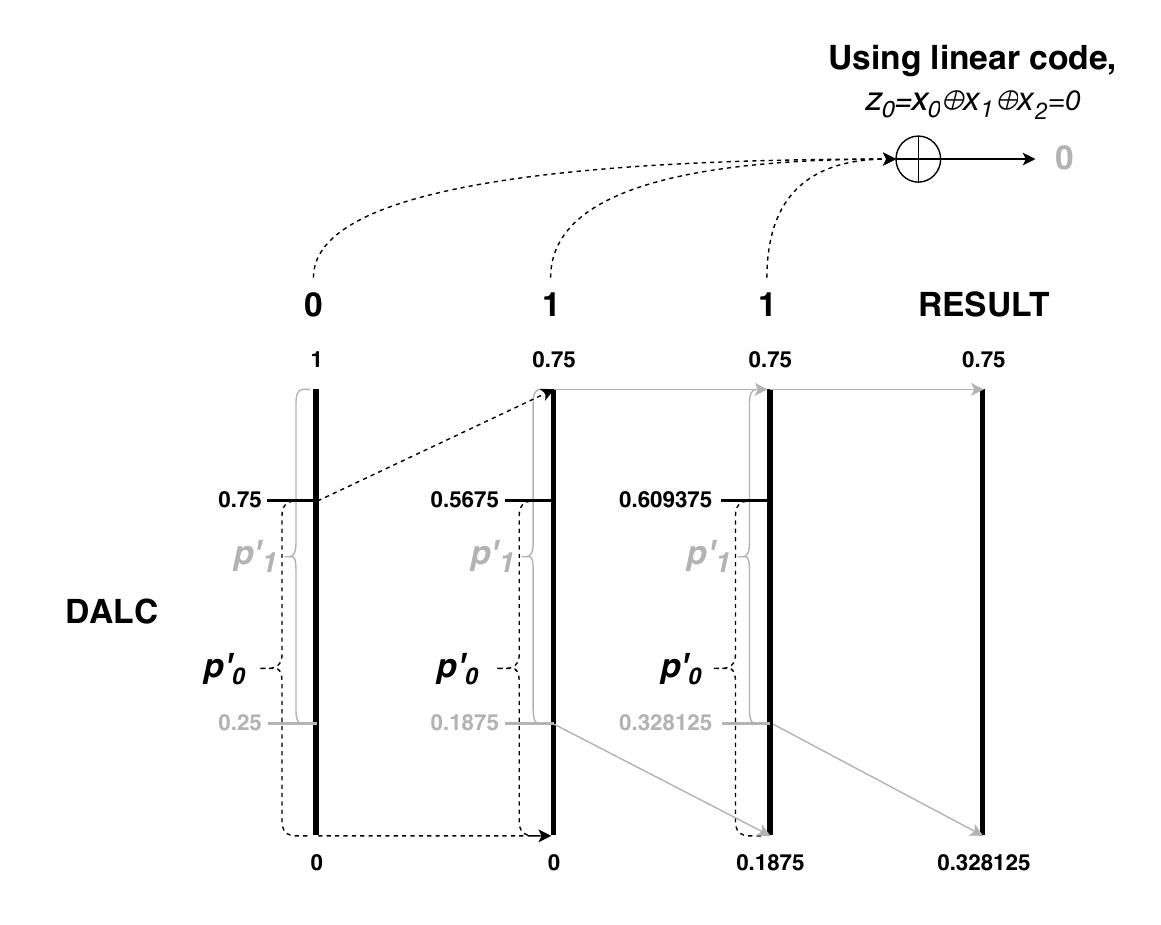}
  \caption{The encoding processes of DALC. The coded intervals of DALC is $[0.328125, 0.75)$. }
  \label{fig:encoding}
\end{figure}

For easy description, we have an example of the encoding process of DALC  illustrated in Fig. \ref{fig:encoding}. Suppose we have $X=011$ as the input. DALC utilizes a linear systematic code coding, where $ z_0 = x_0 \oplus x_1 \oplus x_2 =0$, sets the parity check bit to ``0", thereby generating the parity check sequence $Z=0$.

\subsection{Decoder of DALC}

\begin{algorithm}
\caption{Decoding process of DALC}
\label{alg:example}
\begin{algorithmic}[1]
\Require Codeword $C_{X}$, Length $n$, Termination $t$, The sizes of intervals corresponding to 0 and 1 $p_{0}$ and $p_{1}$, Overlap factor $\alpha$, Generator matrix $\mathbf{G}$, Parity checks sequence $Z$, Maximum number of candidate paths $M$.

\State $i \gets 0$, $(L_{i},H_{i}] \gets (0,1]$.

\State $i \gets 0$, $K \gets 0$,  $(L_{i}^{K},H_{i}^{K}] \gets (0,1]$, $\Omega \gets \emptyset$, $X_{c}^{K} \gets \emptyset$.

\For{$i=1$ {to} $n-1$}
   \If{$i \geq n-t$}
   \State $\alpha \gets 1$
   \EndIf
   
    \If{$L_{i}^{K}<C_{X}<H_{i}^{K}-p_{1}^{\alpha}(H_{i}^{K}-L_{i}^{K})$}
        \State $X_{c}^{K} \gets X_{c}^{K} \cup 0$
        \State $\Omega \gets \Omega \cup X_{c}^{K}$
        \State $(L_{i+1}^{K},H_{i+1}^{K}] \gets (L_{i}^{K},L_{i}^{K}+p_{0}^{\alpha}(H_{i}^{K}-L_{i}^{K})]$
    \ElsIf{$L_{i}^{K}+p_{0}^{\alpha}(H_{i}^{K}-L_{i}^{K})<C_{X}<H_{i}^{K}$}
        \State $X_{c}^{K} \gets X_{c}^{K} \cup 1$
        \State $\Omega \gets \Omega \cup X_{c}^{K}$
        \State $(L_{i+1}^{K},H_{i+1}^{K}] \gets (H_{i}^{K}-p_{1}^{\alpha}(H_{i}^{K}-L_{i}^{K}),H_{i}^{K}]$
    \ElsIf{$L_{i}^{K}+p_{0}^{\alpha}(H_{i}^{K}-L_{i}^{K})<C_{X}<H_{i}^{K}-p_{1}^{\alpha}(H_{i}^{K}-L_{i}^{K})$} 
        \State $X_{c}^{K} \gets X_{c}^{K} \cup 0$
        \State $X_{c}^{K+1} \gets X_{c}^{K} \cup 1$
        \State $(L_{i+1}^{K},H_{i+1}^{K}] \gets (L_{i}^{K},L_{i}^{K}+p_{0}^{\alpha}(H_{i}^{K}-L_{i}^{K})]$
        \State $(L_{i+1}^{K+1},H_{i+1}^{K+1}] \gets (H_{i}^{K}-p_{1}^{\alpha}(H_{i}^{K}-L_{i}^{K}),H_{i}^{K}]$
        \State $\Omega \gets \Omega \cup X_{c}^{K} \cup X_{c}^{K+1}$
        \State $K \gets K+1$
    \EndIf
    
    \State Sort the candidate paths $X_{c}$ within $\Omega$ in descending order based on their $\hat{X}_{MAP}$, which is computed by equation (2).

    \If{$K>M$}
    \State $K\gets M$
    \State Retain the top $M$ candidate paths within $\Omega$
    \EndIf

\EndFor

\For{ $j=0$ {to} $K$ }
    \State $Z_{c}^j \gets X_{c}^{j}\mathbf{G}^T$
    \If{$Z_{c}^j==Z$}
    \Return $X_{c}^j$
    \EndIf
\EndFor

\Return False

\end{algorithmic}
\end{algorithm}

\begin{figure*}[ht]
  \centering
  \includegraphics[width=16cm, page={1}]{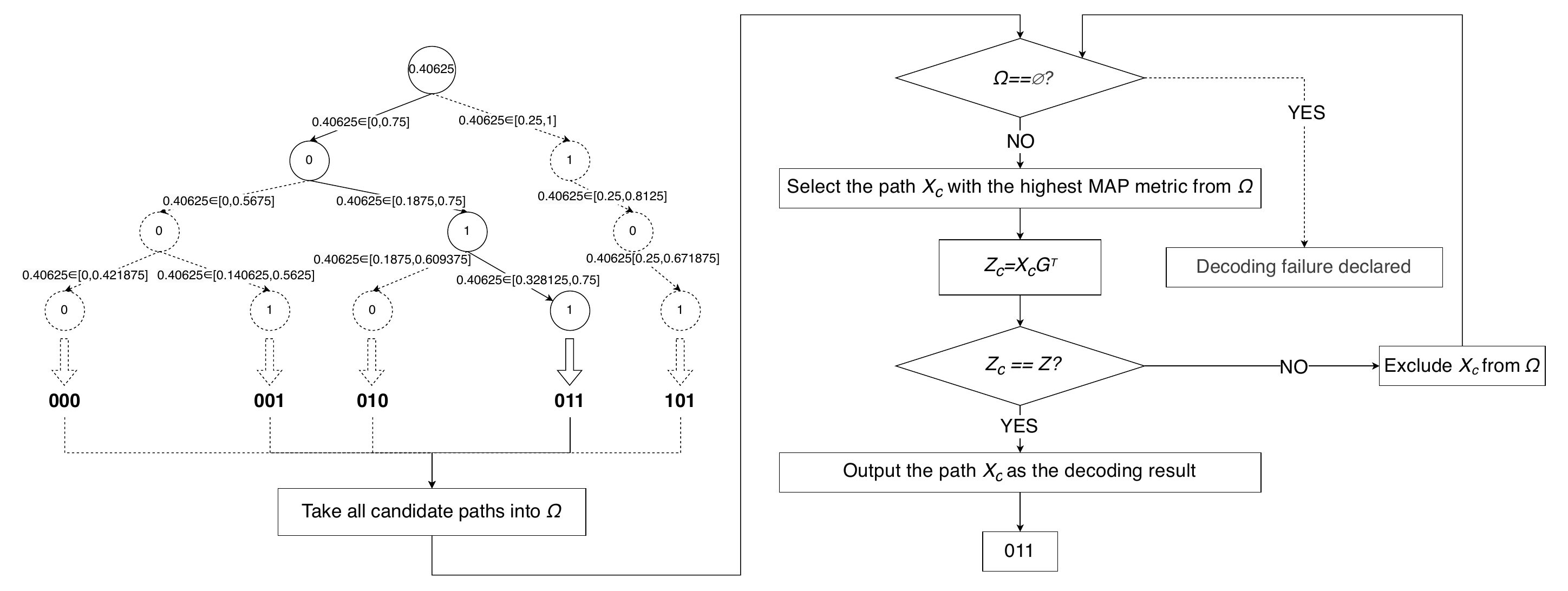}
  \caption{A decoding example of DALC. The decoding result of DALC is the correct path ``011". }
  \label{fig:decoding}
\end{figure*}

The DALC decoding example is illustrated in Fig. \ref{fig:decoding}. Within the output interval presented in Fig. \ref{fig:encoding}, we selected ``0.40625" as the decoding input, which can be represented as the binary number ``0.01101". Fig. \ref{fig:decoding} shows that the DALC decoding result correctly outputs the decoding path ``011". During the Path Selection phase, the candidate path ``001," which had a higher MAP metric, did not satisfy the linear code verification and was consequently eliminated.

\section{Experiment and Result}
\label{result}

This section compares the coding performance of DALC, MDAC \cite{MDAC2020}, ADAC \cite{ADAC2023}, and DAC \cite{DAC2009}. It also discusses the impact of linear codes' error-detection capability and their specific implementation. The experimental setup follows \cite{MDAC2020}, with M=2048 candidate paths and a source sequence of length n=600. The BER is obtained by varying the crossover probability for different entropy values H(X,Y). Results include balanced ($p_0 = 0.5$) and skewed ($p_0 = 0.3, 0.1$) scenarios, with each data point averaging $10^3$ trials, as shown in Figures \ref{fig:0.5}, \ref{fig:0.3} and \ref{fig:0.1}.

Evidence from \cite{DAC2009} suggests that mapping all bits of the source onto partially overlapped intervals can significantly degrade the performance of DAC. Consequently, for a source of length $n$, only the first $n - t$ bits are mapped onto partially overlapped intervals, while the remaining $t$ bits at the tail are encoded using classic AC \cite{DAC2009}. In the experiments, we also compared the decoding performance with termination lengths $t= 4$.

\subsection{Specific Implementation Of Linear Codes}
\subsubsection{Cyclic Redundancy Check (CRC)}
In this paper, we utilize the CRC16 with the generator polynomial $G_{CRC}(x) = x^{16} + x^{15} + x^{13} + x^0$ . Taking the binary sequence ``1010" as an example, the parity checks sequence $Z$ is ``1000000000001100".

\subsubsection{Bose-Chaudhuri-Hocquenghem (BCH)}
In this paper, we establish the BCH code with a design distance $d = 5$, and its generator polynomial $G_{BCH}(x) = x^{15} + x^{13} + x^{10} + x^{6} + x^{4} + x^{2} + x^{0}$. Using the binary sequence ``1010" as an example, the sequence with the appended BCH code is ``1011001111111100".

\subsubsection{Low-Density Parity Check (LDPC)}
In this paper, the LDPC codes we employ are constructed in the following manner. The source sequence is partitioned into 16 groups in an interleaved manner: within every set of 16 consecutive bits, the first bit is assigned to the first group, the second bit to the second group, and so on, until the 16-th bit is assigned to the 16-th group. Each group computes a single parity bit based on its assigned bits. For example, given the binary source sequence "1010001010100010", the generated error correction code $Z$ is "1010001010100010", where each bit corresponds to the parity bit computed by the respective group. This interleaved grouping ensures robust error detection and correction, especially in scenarios with burst errors.

\subsubsection{Block Parity Check (BPC)}
In this paper, the BPC codes we employ are constructed in the following manner. The source sequence is divided into 16 contiguous segments, where each segment consists of $n/16$ consecutive bits. Each segment independently computes a single parity bit based on its assigned bits. Unlike LDPC, the grouping here is non-interleaved, and the parity bits are computed directly from contiguous blocks of the source sequence. For example, given the binary sequence "1010001010100010", the sequence after incorporating BPC codes becomes "1010001010100010", where the parity bits are derived from the contiguous segments. BPC is particularly suited for scenarios requiring low computational complexity and basic error detection.

\subsection{Comparison Of Decoding Performance}

%

\begin{figure*}
\centering
\begin{minipage}{0.5\textwidth}
  \centering
  \includegraphics[width=\linewidth]{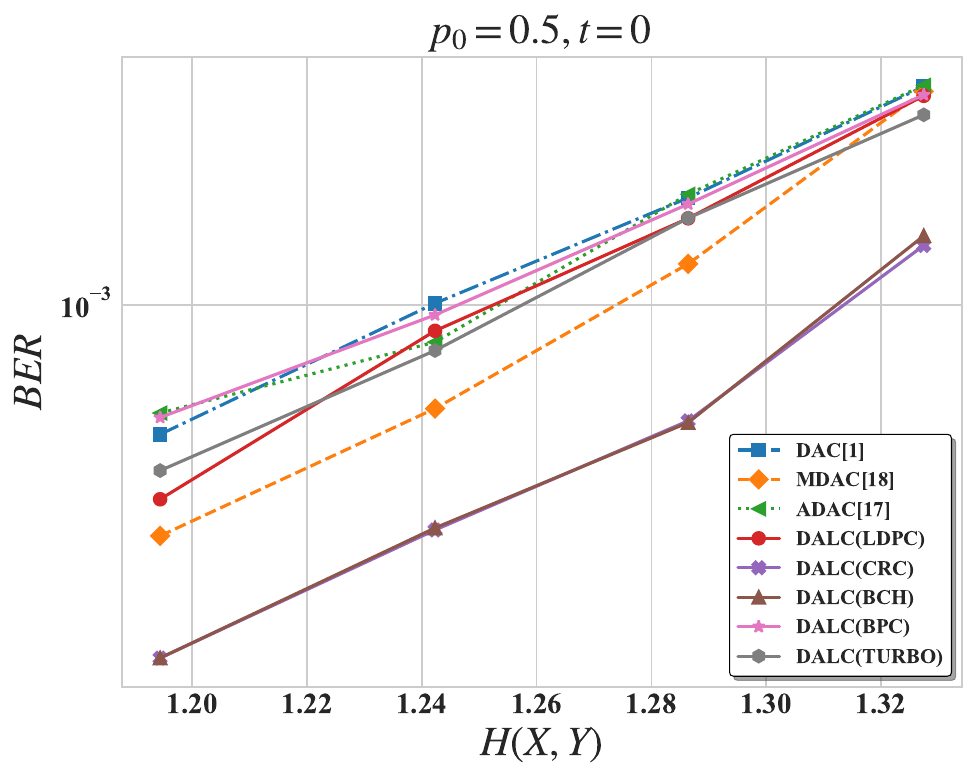}
  \par (a)
\end{minipage}%
\begin{minipage}{0.5\textwidth}
  \centering
  \includegraphics[width=\linewidth]{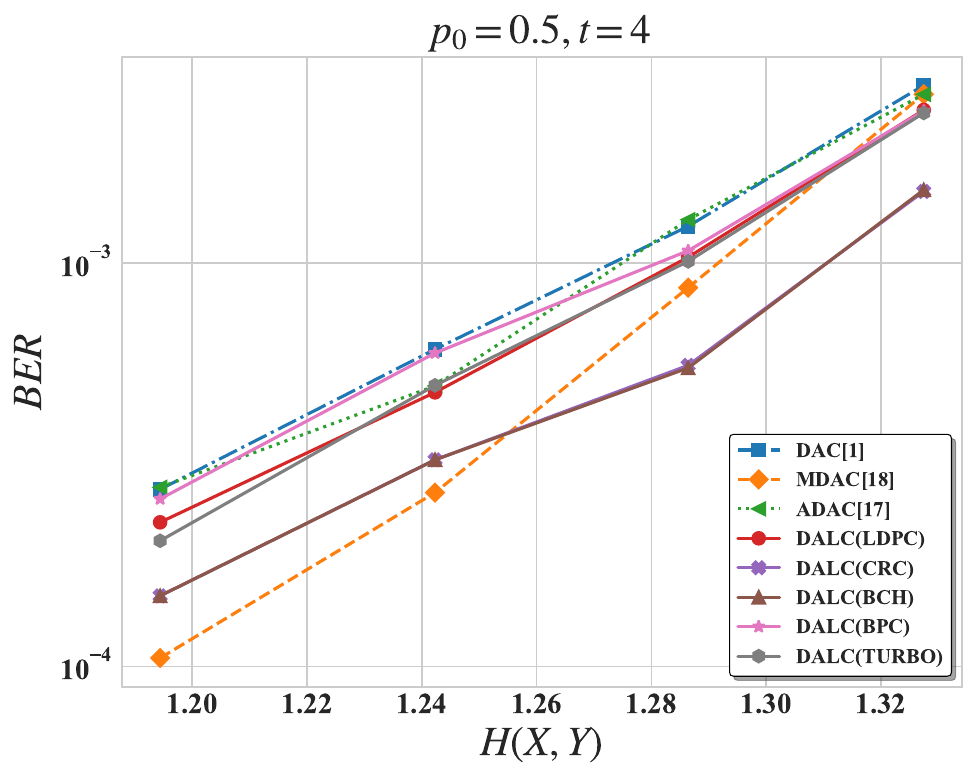}
  \par (b) 
\end{minipage}%
\caption{Comparison of decoding performance with $p_0=0.5$, two scenarios are considered: 
 (a) $t=0$; (b) $t=14$. }
\label{fig:0.5}
\end{figure*}

\begin{figure*}
\centering
\begin{minipage}{0.5\textwidth}
  \centering
  \includegraphics[width=\linewidth]{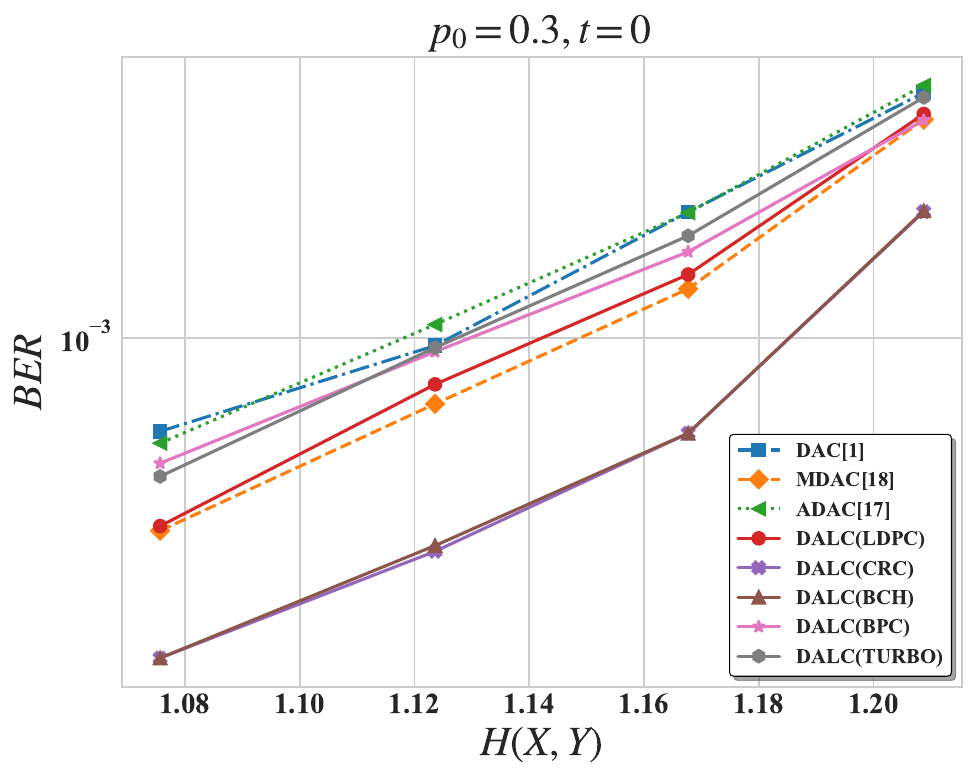}
  \par (a)
\end{minipage}%
\begin{minipage}{0.5\textwidth}
  \centering
  \includegraphics[width=\linewidth]{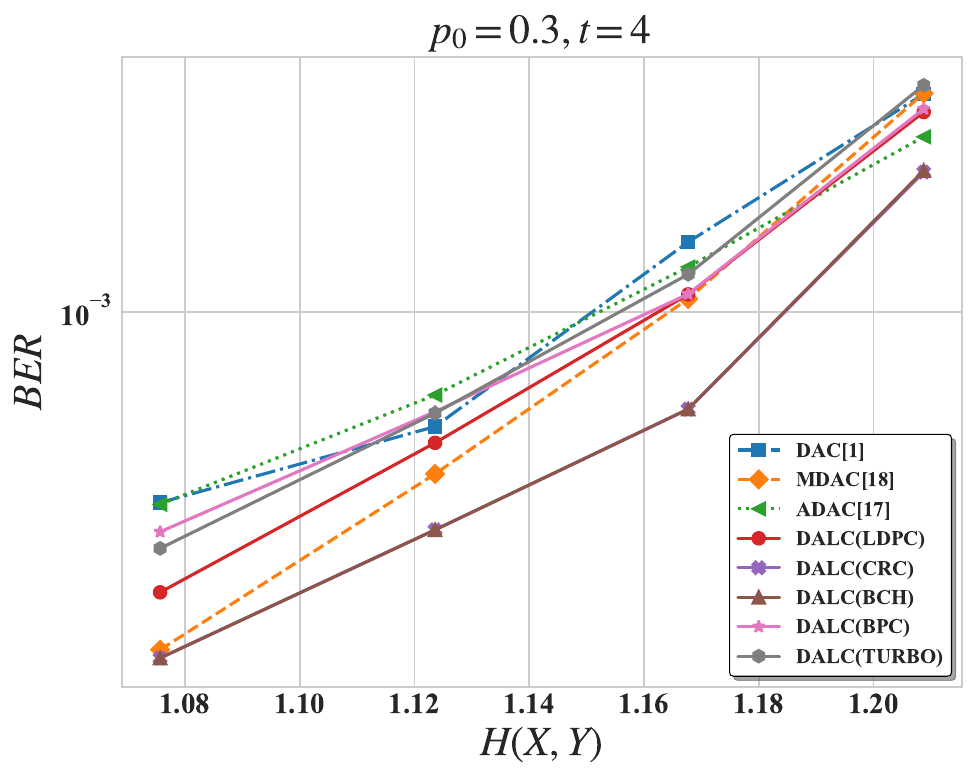}
  \par (b) 
\end{minipage}%
\caption{Comparison of decoding performance with $p_0=0.3$, two scenarios are considered: 
 (a) $t=0$; (b) $t=4$. }
\label{fig:0.3}
\end{figure*}

\begin{figure*}
\centering
\begin{minipage}{0.5\textwidth}
  \centering
  \includegraphics[width=\linewidth]{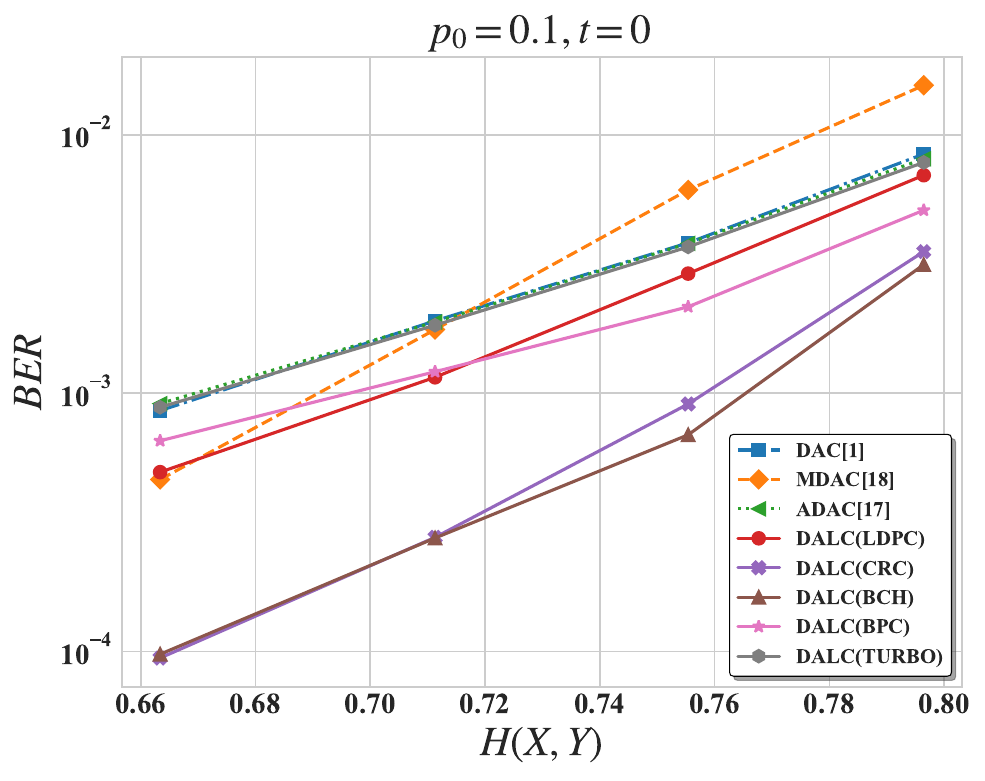}
  \par (a)
\end{minipage}%
\begin{minipage}{0.5\textwidth}
  \centering
  \includegraphics[width=\linewidth]{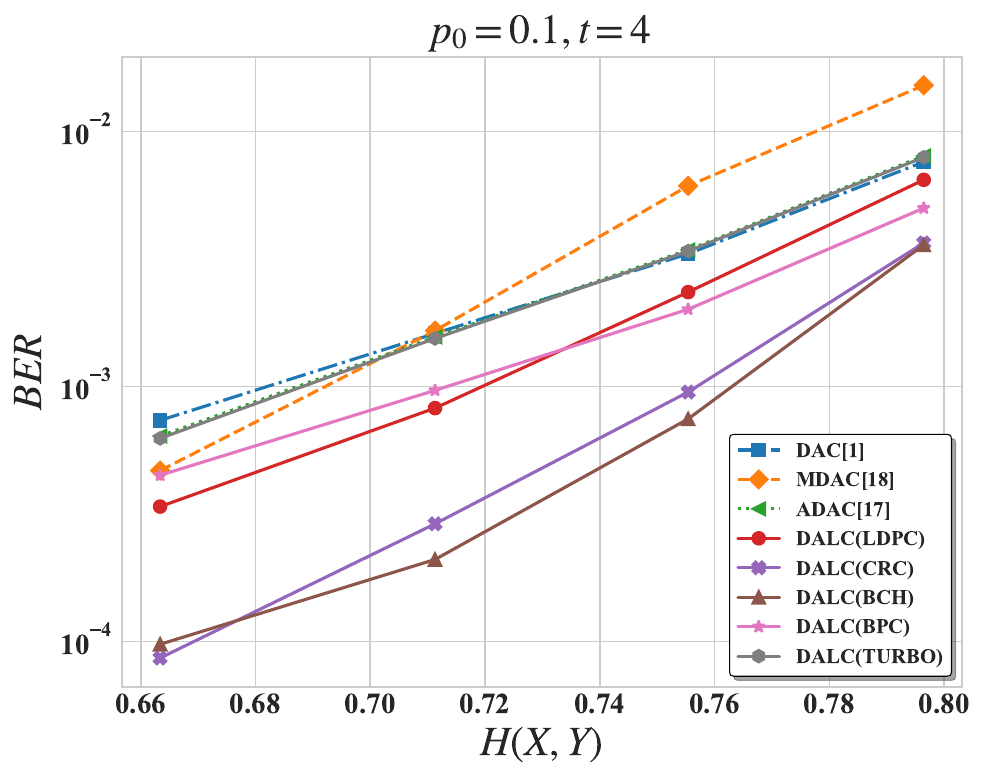}
  \par (b) 
\end{minipage}%
\caption{Comparison of decoding performance with $p_0=0.1$, two scenarios are considered: 
 (a) $t=0$; (b) $t=4$. }
\label{fig:0.1}
\end{figure*}

From Fig. \ref{fig:0.5}(a),\ref{fig:0.3}(a) and \ref{fig:0.1}(a), it is evident that the DALC with different linear codes, particularly CRC and BCH, outperform other methods across the tested range of $H(X, Y)$ when $t=0$. This performance improvement is attributed to the integration of linear codes during the encoding phase and the execution of linear code verification during the decoding phase, which reduces the number of candidate paths and eliminates some interfering paths, thereby lowering the BER.

When  $p_0 = 0.5$, as depicted in Fig. \ref{fig:0.5}, the DALC employing BCH and CRC methods exhibits a significantly lower BER compared to MDAC\cite{MDAC2020} and ADAC\cite{ADAC2023}. Furthermore, under the conditions of $t = 0$ and $t = 4$, as shown in Fig. \ref{fig:0.5}(a) and Fig. \ref{fig:0.5}(b), the performance advantage is less pronounced when $t = 4$.

When the skewness of the source increases, i.e., at  $p_0 = 0.3$, as shown in Fig. \ref{fig:0.3}, the performance gap between the DALC methods and MDAC\cite{MDAC2020} and ADAC\cite{ADAC2023} diminishes, but the DALC methods employing BCH and CRC still maintain a lower BER. 

When the source skewness increases to $p_0 = 0.1$, as depicted in Fig. \ref{fig:0.1}, DALC methods outperform MDAC\cite{MDAC2020} and ADAC\cite{ADAC2023}, demonstrating their robustness with imbalanced data.

As shown in Fig. \ref{fig:0.5}(a), Fig. \ref{fig:0.3}(a), and Fig. \ref{fig:0.1}(a), the decoding performance assisted by CRC and BCH outperforms that of BPC or LDPC under the condition of $t=0$. This can be attributed to the fact that both CRC and BCH are generated based on polynomial division over Galois fields, which results in a lower undetected error rate compared to BPC and LDPC, which rely on modulo-2 addition. The algebraic structure of CRC and BCH enables more robust error detection and correction, particularly in scenarios where precise error localization is critical. Additionally, due to the inherent complexity of the decoding process, LDPC and BPC may struggle to effectively handle consecutive errors at the tail of the decoding path, where error propagation can significantly degrade performance. This limitation highlights the advantage of CRC and BCH in managing tail-end errors, making them more suitable for applications requiring high reliability in distributed arithmetic coding frameworks.

As illustrated in Fig. \ref{fig:0.5}(a) and Fig. \ref{fig:0.5}(b), Fig. \ref{fig:0.3}(a) and Fig. \ref{fig:0.3}(b), and Fig. \ref{fig:0.1}(a) and Fig. \ref{fig:0.1}(b), the performance advantage of DALC under the condition of $t=4$ is less significant compared to the case of $t=0$. This can be attributed to the strategy of appending $t$ overlapping bits with a factor of 1 at the tail, which effectively reduces ambiguity during the decoding of the tail section and mitigates the issue of errors concentrated at the tail of the decoding path. However, this approach partially overlaps with one of the key benefits of linear code-assisted decoding, namely, the ability to constrain and correct errors in the decoding process. As a result, the performance improvement of DALC is less pronounced when $t=4$ than when $t=0$.

\begin{figure}[ht]
  \centering
  \includegraphics[width=8cm, page={1}]{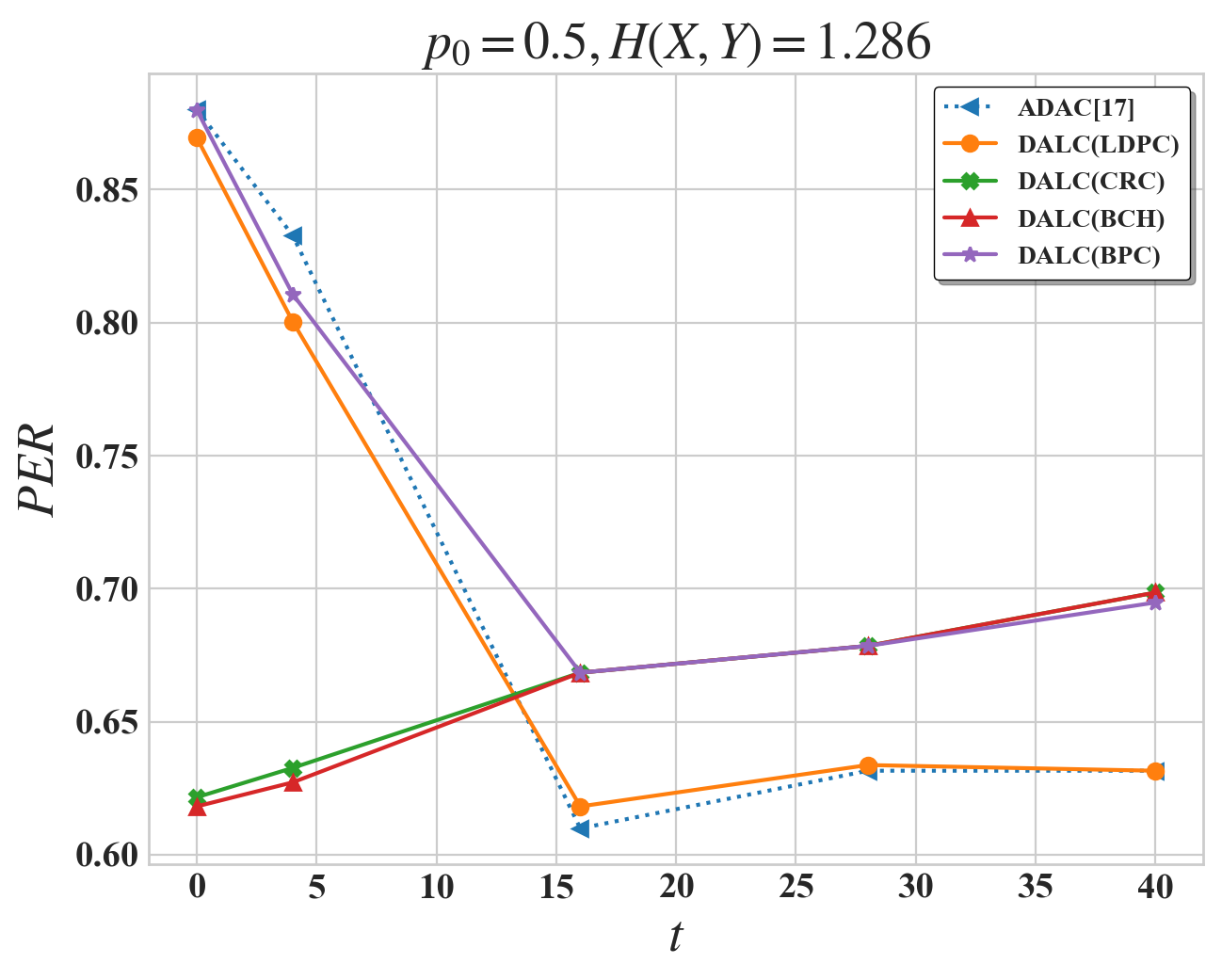}
  \caption{Proportion of erroneous bits located in the latter half of the block.}
  \label{fig:tailerror}
\end{figure}

For DAC, one notable issue is the difficulty in effectively identifying errors occurring in the tail portion. While both linear code screening and setting parameter $t$ can partially address this problem, their individual effects are similar, and combining them leads to suboptimal performance. We analyzed the proportion of erroneous bits located in the latter half of the block, as shown in Fig. \ref{fig:tailerror}. As $t$ increases, the proportion of decoding processes assisted by CRC in DALC exhibits an upward trend, whereas the proportion without CRC assistance shows a downward trend. This indicates that CRC-aided decoding can mitigate some tail-bit errors, and setting $t$ also contributes partially to solving this issue. However, when both strategies are applied simultaneously, they interfere with each other, diminishing their combined effectiveness.

\section{Conclusion}
\label{conclusion}
In this paper, we propose a distributed source coding method known as Distributed Arithmetic Coding Aided by Linear Codes (DALC), which optimizes the decoding process by incorporating linear code constraints within the decoding procedure of Distributed Arithmetic Coding (DAC). By leveraging the constraints of linear codes, we have further regulated the conditions for decoding to conform, excluded candidate paths that do not meet the linear code verification, and improved the decoding performance. Simulation results indicate that DALC with CRC and LDPC can improve decoding performance in terms of BER. Future work will focus on optimizing this integration to enhance decoding accuracy and maintain a high compression ratio. 

\section*{Acknowledgment}

This work was partially supported by the National Key Research and Development Program (Grant Nos. 2022YFB3104001, 2022YFC3321102), the National Natural Science Foundation of China (Grant Nos. 61672398, 61806151). 

\bibliographystyle{IEEEtran}
\bibliography{refernce.bib}

\end{document}